\documentclass[epj]{svjour}
\usepackage{graphics}
\usepackage{graphicx}
\usepackage{subfigure}
\usepackage{amssymb,amsmath}
\usepackage{indentfirst}
\usepackage{amsmath}
\usepackage[square, comma, sort&compress, numbers]{natbib}
\newcommand{\beq}{\begin{equation}}
\newcommand{\eeq}{\end{equation}}

\def\ba{\begin{eqnarray}}
\def\ea{\end{eqnarray}}

\begin{document}

\authorrunning{Ma, et al.}
\titlerunning{Implications of the lens redshift distribution of strong lensing systems}

\title{Implications of the lens redshift distribution of strong lensing systems: cosmological parameters and the global properties of
early-type galaxies}
\author{Yu-Bo Ma\inst{1}, Shuo Cao\inst{2}\thanks{\emph{e-mail:}
caoshuo@bnu.edu.cn}, Jia Zhang \inst{3}, Shuaibo
Geng\inst{2}, Yuting Liu\inst{2}, Tonghua Liu\inst{2}, and Yu Pan\inst{4} 
%
}                     
%
%
\institute{Department of Physics, Shanxi Datong University, Datong,
037009, China; \and Department of Astronomy, Beijing Normal
University, Beijing, 100875, China; \and Department of Physics,
School of Mathematics and physics, Weinan Normal University, Shanxi
714099, China; \and College of Science, Chongqing University of
Posts and Telecommunications, Chongqing, 400065, China}
%
%
\abstract{In this paper, we assemble a well-defined sample of
early-type gravitational lenses extracted from a large collection of
158 systems, and use the redshift distribution of galactic-scale
lenses to test the standard cosmological model ($\Lambda$CDM) and
the modified gravity theory (DGP). Two additional sub-samples are
also included to account for possible selection effect introduced by
the detectability of lens galaxies. Our results show that
independent measurement of the matter density parameter ($\Omega_m$)
could be expected from such strong lensing statistics. Based on
future measurements of strong lensing systems from the forthcoming
LSST survey, one can expect $\Omega_m$ to be estimated at the
precision of $\Delta\Omega_m\sim 0.006$, which provides a better
constraint on $\Omega_m$ than \textit{Planck} 2015 results.
Moreover, use the lens redshift test is also used to constrain the
characteristic velocity dispersion of the lensing galaxies, which is
well consistent with that derived from the optical spectroscopic
observations. A parameter $f_E$ is adopted to quantify the relation
between the lensing-based velocity dispersion and the corresponding
stellar value. Finally, the accumulation of detectable galactic
lenses from future LSST survey would lead to more stringent fits of
$\Delta f_E\sim10^{-3}$, which encourages us to test the global
properties of early-type galaxies at much higher accuracy.
%
} 
\maketitle
\section{Introduction}

The current accelerating expansion of the Universe, which is
supported by the observations of Type Ia supernovae (SNIa)
\citep{Riess98,Perlmutter99} in combination with independent
estimates of cosmic microwave background (CMB) \citep{Spe03} and
large scale structure (LSS) \citep{Eisenstein05}, has become one of
the fundamental challenges to standard models in particle physics
and modern cosmology. As was pointed out in an increasing body of
literature \citep{Ellis05,Wiltshire07}, this cosmic acceleration can
be attributed to an energy component with negative pressure (the so
called dark energy), which dominates the universe at late times and
causes the observed accelerating expansion. The other possibility is
to contemplate modifications to the Friedman- Robertson-Walker
models arising from extra dimensions, which has triggered many
theoretical speculations in the so-called brane-world scenarios
\citep{Dvali00}. However, the potential of certain type of
observational data, even though ever increasing, does not yet allow
us to differentiate the two likely explanations for the observed
cosmic acceleration. For instance, the SNIa data would not be
sufficient to place stringent constraint on cosmological parameters,
if taken alone separately from the other approaches
\citep{Hinshaw09}. Indeed, the power of modern cosmology lies in
building up consistency rather than in single and precise
experiments \citep{Biesiada10}, which indicates that every
alternative method of restricting cosmological parameters is
desired. Following this direction, a number of combined analyses
involving baryonic fraction the x-ray gas mass fraction of clusters
\citep{Allen08}, radio observations of the Sunyaev-Zeldovich effect
together with X-ray emission \citep{De05}, and ultra-compact
structure in intermediate-luminosity radio quasars
\citep{Cao17a,Cao17b} have been performed in the literature, which
were able to constrain the cosmological parameters consistent with
the analysis of Type Ia supernovae. In this paper, we will assemble
a large sample of strongly gravitationally lenseing systems (SGL)
\citep{Cao15,Shu17} to examine whether the lens redshift
distribution test can be utilized - not only to optimize the
parameters in $\Lambda$CDM model - but also to carry out comparative
studies between competing cosmologies.

In the past decades, an cosmological examination of galactic-scale
SGL systems, based on the derived angular diameter distances between
the source, the lens, and the observer, has been applied to test a
diverse range of dynamical dark energy models. For example, the XCDM
model in which dark energy is described by a hydrodynamic
energy-momentum tensor with a constant EoS coefficient, and the
holographic dark energy model arising from the holographic principle
of quantum gravity theory \citep{Li16}. On the other hand, the first
attempt to determine cosmological parameters from the redshift
distribution of the lensing galaxies was presented in
\citet{Ofek03}, which investigated the viability of using lens
redshift test to place additional constraints on dark energy models.
The original (to our knowledge) formulations of this approach can be
traced back to \citet{Kochanek92,Helbig96,Kochanek96a}. The purpose
of this paper is to extend our previous statistical analysis based
on the angular separation distribution of the lensed images
\citep{Cao11c,Cao12a}, and show how the lens redshift distribution
of the most recent and significantly improved observations of
early-type gravitational lenses (158 combined systems) can be used
to provide accurate estimates of the cosmological constant in the
$\Lambda$CDM model and the parameters of alternative cosmological
models.

More importantly, in the framework of the concordance cosmological
model ($\Lambda$CDM) and a singular isothermal ellipsoid (SIE) model
for galactic potentials, strong lensing statistics have been most
often used for a different purpose: to study the number density of
lensing galaxies as a function of redshift, i.e., the velocity
dispersion function (VDF) of potential lenses, since strong lensing
probability is proportional to the comoving number density times
$\sigma^4$ (the image separation is proportional to $\sigma^2$)
\citep{Chae07}. Meanwhile, considering the fact that early-type
galaxies dominate the lensing cross sections due to their larger
central mass concentrations, gravitational lenses therefore provide
a unique mass-selected sample to study the global properties of
early-type galaxies over a range of redshifts (up to $z\sim1$). A
pioneer work was made by \citet{Chae05}, which investigated the VDF
of early-type galaxies based on the distribution of lensed image
separations observed in the Cosmic Lens All-Sky Survey (CLASS) and
the PMN-NVSS Extragalactic Lens Survey (PANELS). However, the
statistical lens sample of lensed systems, which are required to be
complete for image separations, is too small to provide accurate
estimates. In this work, focusing a larger sample of 158
gravitational lenses drawn from the Sloan Lens ACS (SLACS) Survey
and other sky surveys \citep{Cao15,Shu17}, we will use the
distribution of lens reshifts to provide independent constraints on
the velocity dispersion function of early-type galaxies
($z\sim1.0$), especially the characteristic velocity dispersion
($\sigma_*$) in a solely lensing-based VDF.

This paper is organized as follows. In Section 2 we briefly describe
the methodology and the lens redshift data from various surveys. In
Section 3 we introduce two prevalent cosmologies and show the
fitting results on the relevant cosmological parameters. In Section
4 we present the constraints on a model VDF of early-type galaxies
and discuss their implications. The final conclusions are summarized
in Section 5.


\section{Methodology and observations}
\label{sec:method}

On the assumption that early-type galaxies are uniformly distributed
in comoving space and distributed in luminosity following a
Schechter (1976) luminosity function, the differential probability
of the ray from the background source encountering a lens per unit
redshift is \citep{Turner84}
\begin{equation}
\frac{d\tau}{dz_l}=n(\theta_E,z_l)(1+z_l)^{3} S_\mathrm{cr}
\frac{cdt}{dz_l} \, , \label{opt_depth}
\end{equation}
where $n(\theta,z_l)$ is the number density of the lenses, $S_{cr}$
is the lensing cross-section for multiple imaging with Einstein
radius $\theta_E$. The proper distance interval $c {\rm d}t/{\rm
d}z_l$ in the FLRW metric is calculated to be
\begin{equation}
\frac{cdt}{dz_l}=\frac{c}{(1+z_l)}\frac{1}{H(z_l; \textbf{p})}
\end{equation}
where $c$ is the speed of light, $H(z_l)$ is the Hubble parameter
(the expansion rate of the Universe) at redshift $z_l$, which is
also dependent on the cosmological parameters \textbf{p}.

First of all, concerning the radial mass distribution of early-type
galaxies, we use the spherically symmetric power-law mass
distribution as the lens model, which has been extensively used in
recent studies of lensing caused by early-type galaxies
\citep{Treu06,Cao15,Li16}. In the framework of a general mass model
for the total (i.e. luminous plus dark-matter) mass density ($\rho$)
and luminosity density ($\nu$) \citep{Koopmans05}
\begin{eqnarray}
\label{eq:rhopl}
\rho(r) &=& \rho_0 \left(\frac{r}{r_0}\right)^{-\gamma} \nonumber \\
\nu(r) &=& \nu_0 \left(\frac{r}{r_0}\right)^{-\delta}
\end{eqnarray}
$\gamma=\delta$ denotes that the shape of the luminosity density
follows that of the total mass density, while $\gamma=2$ describes a
sphere of collisional ideal gas in equilibrium between thermal
pressure and self gravity. Note that the measurement of Einstein
radius ($\theta_E$) provides us with the mass $M_{lens}$ inside
$\theta_E$:
\begin{equation}
M_{lens} = \pi \theta_E^2 D_l^2 \Sigma_{cr}
\end{equation}
where $\Sigma_{cr}$ denotes the critical projected mass density
$\Sigma_{cr}= \frac{c^2}{4 \pi G} \frac{D_s}{D_l D_{ls}}$ and the
Einstein radius is defined as the radius within which the mean
convergence $\kappa=\Sigma_{E}/\Sigma_{cr}=1$ \footnote{In the
framework of spherically-symmetric distribution, the dimensionless
surface mass density (convergence) of the lens galaxies can be
written as
$\kappa(\theta)=\frac{3-\gamma}{2}\left(\theta_E/\theta\right)^{\gamma-1}$,
where $\theta$ is the angular radius projected to lens plane
\citep{Suyu13}.}. After solving the spherical Jeans equation  based
on the assumption that stellar and mass distributions follow the
same power law, the dynamical mass inside the aperture projected to
lens plane and then scaled to the Einstein radius can be obtained as
\citep{Koopmans05}
\begin{eqnarray} \label{dynamical mass}
M_{dyn}& =&  \frac{\pi}{G} \sigma_{ap}^2 R_E \left( \frac{R_E}{R_{ap}} \right)^{2-\gamma} f(\gamma)\nonumber\\
       &= &\frac{\pi}{G} \sigma_{ap}^2 D_l \theta_E \left( \frac{\theta_E}{\theta_{ap}} \right)^{2-\gamma} f(\gamma)
\end{eqnarray}
The combination of the mass $M_E$ and $M_{dyn}$ will lead to the
following expression \citep{Koopmans05}
\begin{equation} \label{Einstein} \theta_E =   4 \pi
\left(\frac{\sigma_{ap}}{c}\right)^2 \frac{D_{ls}}{D_s} \left(
\frac{\theta_E}{\theta_{ap}} \right)^{2-\gamma} f(\gamma)
\end{equation}
where
\begin{eqnarray} \label{f factor}
f(\gamma) &=& - \frac{1}{\sqrt{\pi}} \frac{(5-2 \gamma)(1-\gamma)}{3-\gamma} \frac{\Gamma(\gamma - 1)}{\Gamma(\gamma - 3/2)}\nonumber\\
          &\times & \left[ \frac{\Gamma(\gamma/2 - 1/2)}{\Gamma(\gamma / 2)} \right]^2
\end{eqnarray}
Note that when $\gamma=2$, the power-law profile will reduce to the
well-known singular isothermal sphere (SIS) model, with the
corresponding Einstein radius as $\theta_E = 4 \pi
\left(\sigma_{SIS}/c\right)^2 D_{ls}/D_s$ \citep{Cao12b,Cao15}. In
our fiducial model, the average logaritmic density slope is modeled
as the results from 58 SLACS strong-lens early-type galaxies with
direct total-mass and stellar-velocity dispersion measurements
\citep{Koopmans09}. $\sigma_{ap}$ is the luminosity averaged
line-of-sight velocity dispersion of the lens inside the aperture
$\theta_{ap}$. More importantly, for a single system the velocity
dispersion is measured within an aperture and then transformed to
that within a circular aperture of radius $R_{eff}/2$ (half the
effective radius). Following the prescription of
\citet{Jorgensen95a,Jorgensen95b}, we transform the velocity
dispersion measured within $R_{eff}/2$:
\begin{equation}
\sigma = \sigma_{ap} (\theta_{eff}/(2 \theta_{ap}))^{-0.04}.
\end{equation} Finally, $D_{l}$, $D_{s}$, and $D_{ls}$ respectively denote the
angular diameter distances between the observer and the lens, the
observer and the source, the lens and the source. Note that the
cosmological model directly enters through these angular diameter
distances, which, under a Friedman-Walker metric with null space
curvature express as
\begin{eqnarray}
\label{inted} D_A(z_1, z_2;\textbf{p})=\frac{c}{
(1+z_2)}\int_{z_1}^{z_2} \frac{dz'}{H(z';\textbf{p})} \, ,
\end{eqnarray}
The strong lensing cross section is related to the Einstein radius
and the cosmological distances as $S_\mathrm{cr}=\pi (\theta_E
D_l)^2$ \citep{Treu06}.

Secondly, in order to derive the differential lensing probability,
we use the empirically determined velocity-dispersion distribution
function of early-type galaxies. As was pointed out in the previous
analysis of strong lensing statistics \citep{Ofek03,Cao11c}, the
luminosity of a galaxy has a power-law relation to its line-of-sight
velocity dispersion (i.e. the Faber-Jackson relation for early-type
galaxies). Therefore, the Schechter (1976) function is generalized
to be a modified Schechter function \citep{Sheth03}, which may
helpfully describe the number density of galaxies with velocity
dispersion lying between $\sigma$ and $\sigma+d\sigma$: \beq
\label{stat2} \frac{d n}{d \sigma}= n_*\left(
\frac{\sigma}{\sigma_*}\right)^\alpha \exp \left[ -\left(
\frac{\sigma}{\sigma_*}\right)^\beta\right] \frac{\beta}{\Gamma
(\alpha/\beta)} \frac{1}{\sigma} \, , \eeq where $\alpha$ is the
low-velocity power-law index, $\beta$ is the high-velocity
exponential cut-off index, $n_*$ is the integrated number density of
galaxies, and $\sigma_*$ is the characteristic velocity dispersion.
Based on a large sample of galaxies from the SDSS Data Release 5
data set, \citet{Choi07} have measured the VDF for local early-type
galaxies, which became a standard in the studies of gravitational
lensing statistics. However, one should note that for a given galaxy
sample used by \citet{Choi07}, the galaxy number counts start to
become incomplete at low-velocity dispersions, due to the limitation
of absolute magnitude. Therefore, for the parameters $n_*$,
$\sigma_*$, $\alpha$ and $\beta$, we will turn to the early-type VDF
obtained through the powerful Monte Carlo method, based on the
galaxy luminosity functions from the SDSS and intrinsic correlations
between luminosity and velocity dispersion \citep{Chae10} (see
\citet{Biesiada14} for discussion about such choice in view of other
data on velocity dispersion distribution functions). In analogy to
and in order to comply with the previous papers
\citep{Ofek03,Matsumoto08}, we also allow for evolution of the
quantities $n_{*}$ and $\sigma_{*}$, by adopting the power-law
evolution for the number density and the characteristic velocity
dispersion as
\begin{eqnarray}
n_{*}(z_l)&=& n_{*} (1+z_l)^{\nu_n} \\ \nonumber \sigma_{*}(z_l)& =
& \sigma_{*} (1+z_{\rm l})^{\nu_v}
\end{eqnarray}
where $\nu_n$ and $\nu_v$ are constant quantities $\nu_n=\nu_v=0$
corresponds to the no evolution model (see \citet{Ofek03} for more
details). In this paper, we take the parameters of the power-law
evolution model as $(\nu_n, \nu_v)=(-0.23, -0.01)$, which were
predicted by the semi-analytic model after \citet{Kang05,Chae07}.

Following the above mentioned procedure, we can compute the
differential probability of the ray from the background source at
$z_{s}$ with Einstein radius $\theta_E$ encountering a lens per unit
redshift
\begin{equation}
\begin{array}{ll}
\frac{d{\tau}}{dz_l} \left( \theta_E, z_{s} \right) = & \tau_N (1+z_l)^{\left[-\nu_v\alpha/(\gamma-1)+\nu_n \right]} \\
        & \times (1+z_l)^{3} D_{l}^{2} \frac{cdt}{dz_l} \theta_E \left( \frac{\theta_E}{\theta_{E*}} \right)^{\alpha/2} \\
        & \times \exp{[- \left( \frac{\theta_E}{\theta_{E*}} \right)^{\beta/2} (1+z_l)^{-\nu_v\beta/(\gamma-1)} ]} \, ,
\end{array}
\label{dtaudz}
\end{equation}
where the normalization $\tau_{N}=\frac{\pi}{2} n_{*}
\frac{\beta}{\Gamma(\alpha/\beta)}$. In the power-law lens model,
the characteristic Einstein radius is obtained from the combination
of Eq.~(6)-(8):
\begin{equation}
\theta_{E*} = \lambda(e) \left[4\pi
\left(\frac{\sigma_{*}}{c}\right)^2\frac{D_{\rm ls}}{D_{\rm
s}}\left(\frac{\theta_{eff}}{2\theta_{ap}}\right)^{0.08}
\theta_{ap}^{\gamma-2} f(\gamma)\right]^{\frac{1}{\gamma-1}}
\end{equation}
where $\lambda(e)$ is a dynamical normalization factor for
non-spherical galaxies \citep{Keeton98}. For $\lambda(e)$ we assume
the three dimensional shapes of lens galaxies in the combination of
two equal number of extreme cases
\begin{equation}
\lambda(e)=0.5\lambda_{\rm obl}(e)+0.5\lambda_{\rm pro}(e),
\end{equation}
where $\lambda_{\rm obl}(e)$ and $\lambda_{\rm pro}(e)$ respectively
denotes the dynamical normalizations for the oblate and the prolate
isothermal spheroids \citep{Oguri12}
\begin{eqnarray}
\lambda_{\rm obl}(e)&
\approx&\exp\left(0.108\sqrt{e}+0.180e^2+0.797e^5\right),
\\ \nonumber
\lambda_{\rm pro}(e)&\approx& 1-0.258e+0.827e^6. \end{eqnarray} In
our fiducial model, the distribution of the ellipticity is modeled
as a Gaussian distribution with $e=0.25\pm0.2$, which is derived
from the axis ratio distributions of early-type galaxies in the SDSS
survey \citep{Bernardi10}. Finally, we obtain the integrated
probability for the multiple imaging with Einstein radius $\theta_E$
due to early-type galaxies, based on which the relative probability
of finding the lens at redshift $z_l$ for a given lens system is
derived as
\begin{eqnarray}
\delta p_l  =  \frac{d{\tau}}{dz_l}/\tau  =
\frac{d{\tau}}{dz_l}/\int_{0}^{z_s}\frac{d{\tau}}{dz_l}dz_{\rm l} \,
. \label{differential}
\end{eqnarray}

For each systems, we calculate the particular differential
probability $\delta p_l(\bf{p})$ given by Eq.~(\ref{differential})
which is normalized unity. For a statistical sample that contains
$N_l$ multiply imaged sources, the likelihood of the observation
data given the observed lens redshift is calculated as
\begin{equation}
\ln \mathcal{L} = \sum_{l=1}^{N_{\rm{l}}}\ln \delta p_l(\bf{p}),
\label{lnL}
\end{equation}
where $\bf{p}$ denotes the cosmological model parameters (e.g., the
matter density in the Universe $\Omega_{m}$) and the velocity
dispersion function parameters (e.g. the characteristic velocity
dispersion $\sigma_*$). Then one can constrain the model parameters
$\bf{p}$ by minimizing the $\chi^2$  function given by
\begin{equation}
\chi^2 = -2 \ln \mathcal{L} \, . \label{chi}
\end{equation}

Next we will summarize the data both from Sloan Lens ACS Survey
(SLACS) observations and recent large-scale observations of galaxies
that will be used as the input for the statistical lensing model
described above. In order to build an homogeneous galaxy sample, we
limit our analysis to gravitational lenses with early-type
morphology. Compared with the lensing statistics based on the
angular separation distribution of the lensed images \citep{Cao11c},
the advantage of the lens redshift test lies in the fact that a lens
with a large separation will not bias the final results, because
$\theta_E$ is used as prior information in the calculation (see
\citet{Ofek03} for more details). In this paper, we use a combined
sample of $n=158$ strong lensing systems from SLACS (97 lenses taken
from \citet{Bolton08,Auger09,Shu17}), the Strong Lensing Legacy
Survey (SL2S) (31 lenses taken from
\citet{Sonnenfeld13a,Sonnenfeld13b}), the BOSS emission-line lens
survey (BELLS) (25 lenses taken from \citet{Brownstein12}), and
Lenses Structure and Dynamics (LSD) survey (5 lenses from
\citet{Treu02,Koopmans02,Treu04}), which is the largest
gravitational lens sample published in the recent work. This sample
is compiled and summarized in \citet{Cao15,Shu17}, in which all
relevant information necessary to perform statistical analysis (the
redshifts, aperture radius, effective radius, Einstein radius) can
be found. Fig.~1 shows the scatter plot of these lensing systems.

One should note that the 158 lenses used in this work come from
different surveys with vastly different selection functions, which
might affect the resulting redshift distributions of the lenses. For
instance, lenses from SL2S will likely miss lenses with small image
separations of the lensed sources due to the seeing limit of CFHTLS
\citep{Sonnenfeld13a,Sonnenfeld13b}, whereas the SLACS and BELLS
sample will miss the lenses with large image separation due to the
finite Sloan fiber size \citep{Brownstein12,Treu04}. Therefore, in
order to verify the completeness of our final lens sample, we will
use two additional sub-samples to account for the possible selection
functions: 36 lenses from the SL2S and LSD sample (Sample A), and
122 lenses from SLACS and BELLS sample (Sample B).

\begin{figure}
   \centering
   \includegraphics[width=0.5\textwidth]{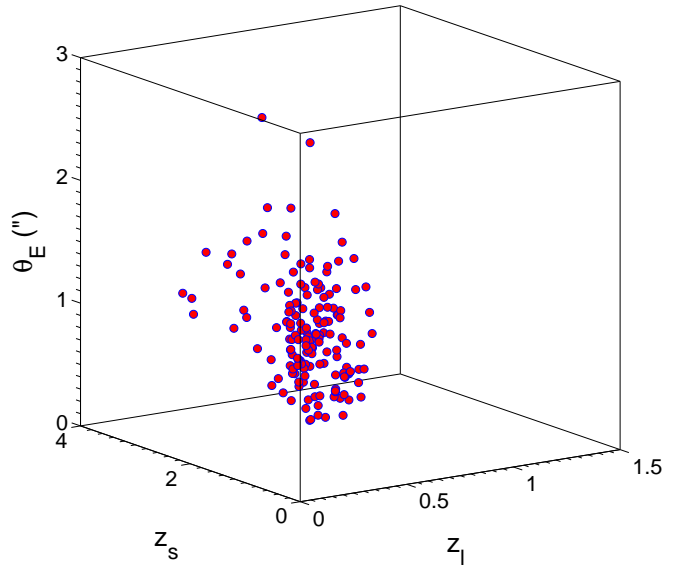}
\caption{ Scatter plot of the current sample of 158 strong lensing
systems. One can see a fair coverage of redshifts in the combined
sample.} \label{fig1}
\end{figure}

\section{Cosmological model and results}
\label{sec:result}

From the observational viewpoint, currently standard cosmological
model, also known as the $\Lambda$CDM model is the simplest one with
constant dark energy density present in the form of cosmological
constant. However, it is important to look into whether the modified
gravity theories are indeed compatible with different kinds of
currently available cosmological data. In this section, we consider
the cosmological constraints on two popular cosmological models, the
$\Lambda$CDM model and the Dvali-Gabadadze-Porrati model arising
from the brane world theory, which have been proposed to explain the
observed cosmic acceleration. For simplicity, a flat Universe is
assumed throughout the following analysis since the spatial
curvature is constrained to be very close to zero with
$|\Omega_k|<0.005$ \citep{Planck15}. Note that the Hubble constant
is not included as a parameter, because the dependence on $H_0$ is
factored out in Eq.~(12). In order to assess the accuracy of our
results, we consider two cases of evolution models of lensing
galaxies: $(\nu_n, \nu_v) = (0, 0)$ and $(\nu_n, \nu_v) = (-0.23,
-0.01)$ \footnote{ We have also performed a sensitivity analysis
through Monte Carlo simulations, in which $\nu_n$ and $\nu_v$ were
respectively characterized by Gaussian distributions with 10\%
uncertainty. The results showed the uncertainties of VDF evolution
parameters have negligible effects on the final cosmological
constraints.}.

\begin{figure*}
   \centering
   \includegraphics[width=0.45\textwidth]{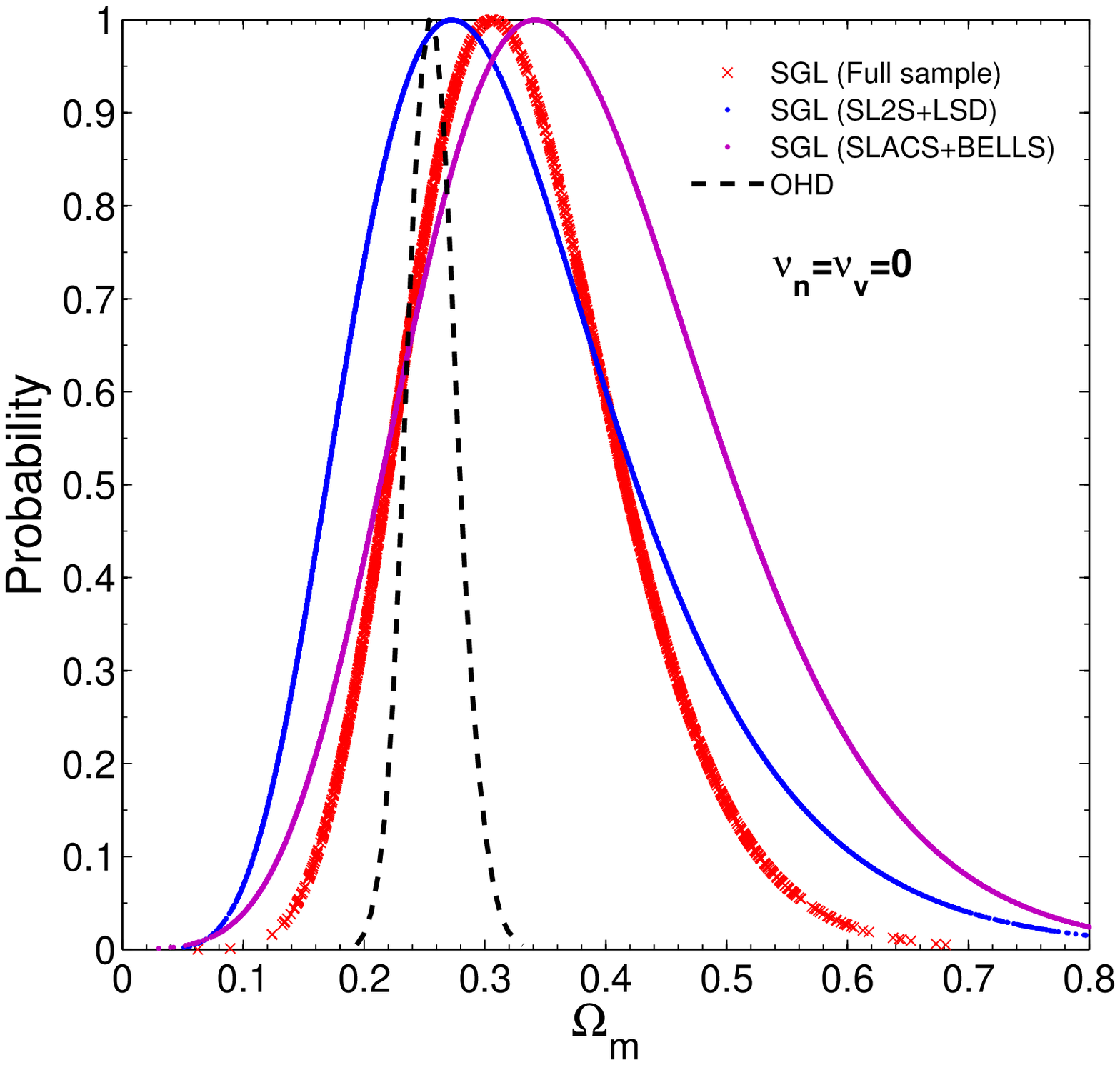}\includegraphics[width=0.45\textwidth]{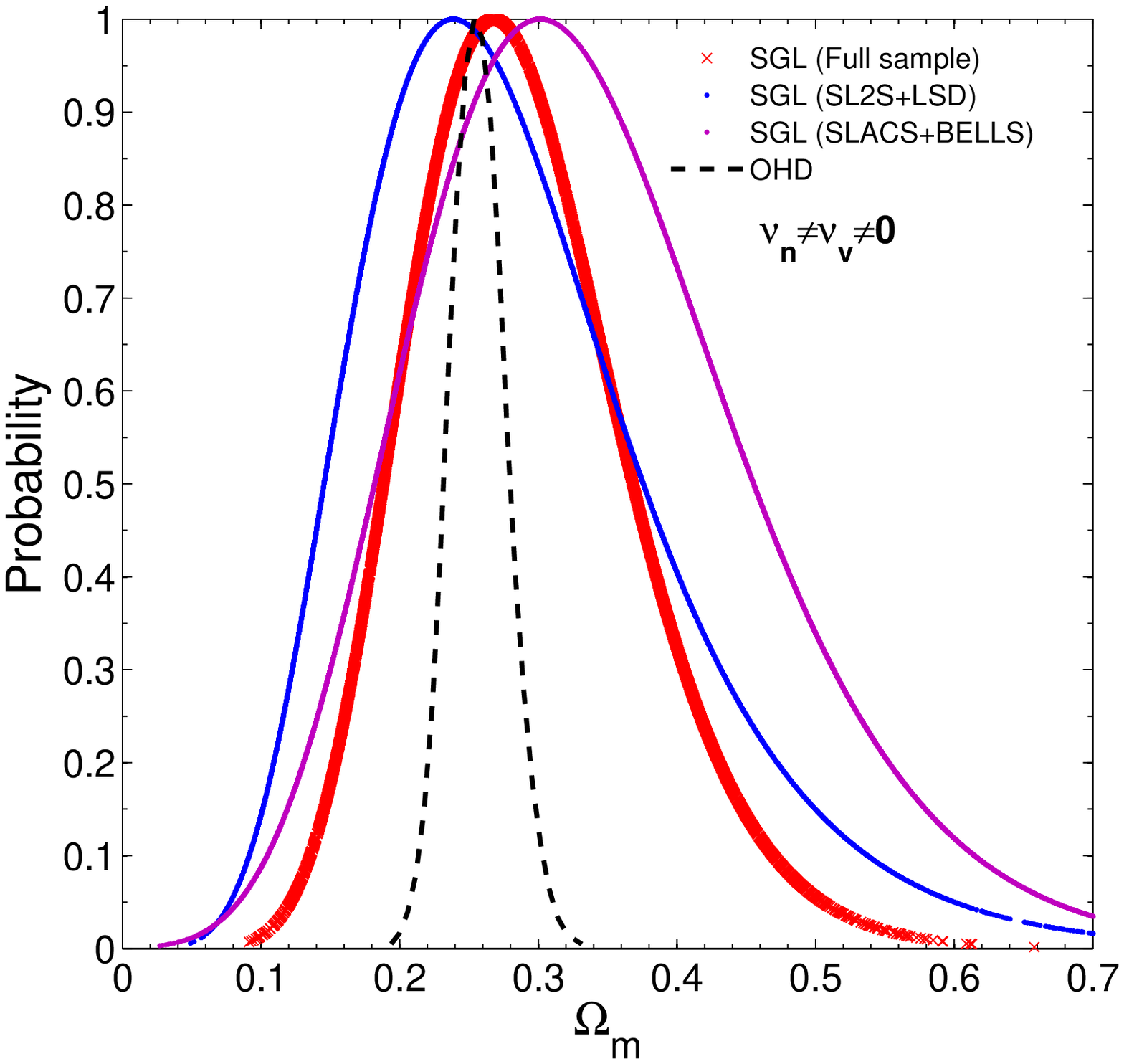}
\caption{ Constraints on the matter density parameter in the flat
$\Lambda$CDM model, which are obtained from the lens redshift
distribution of current SGL systems with and without the redshift
evolution of lensing galaxies. Fitting results from recent
observational Hubble parameter data (OHD) are also added for
comparison.} \label{fig2}
\end{figure*}

\begin{figure*}
   \centering
   \includegraphics[width=0.45\textwidth]{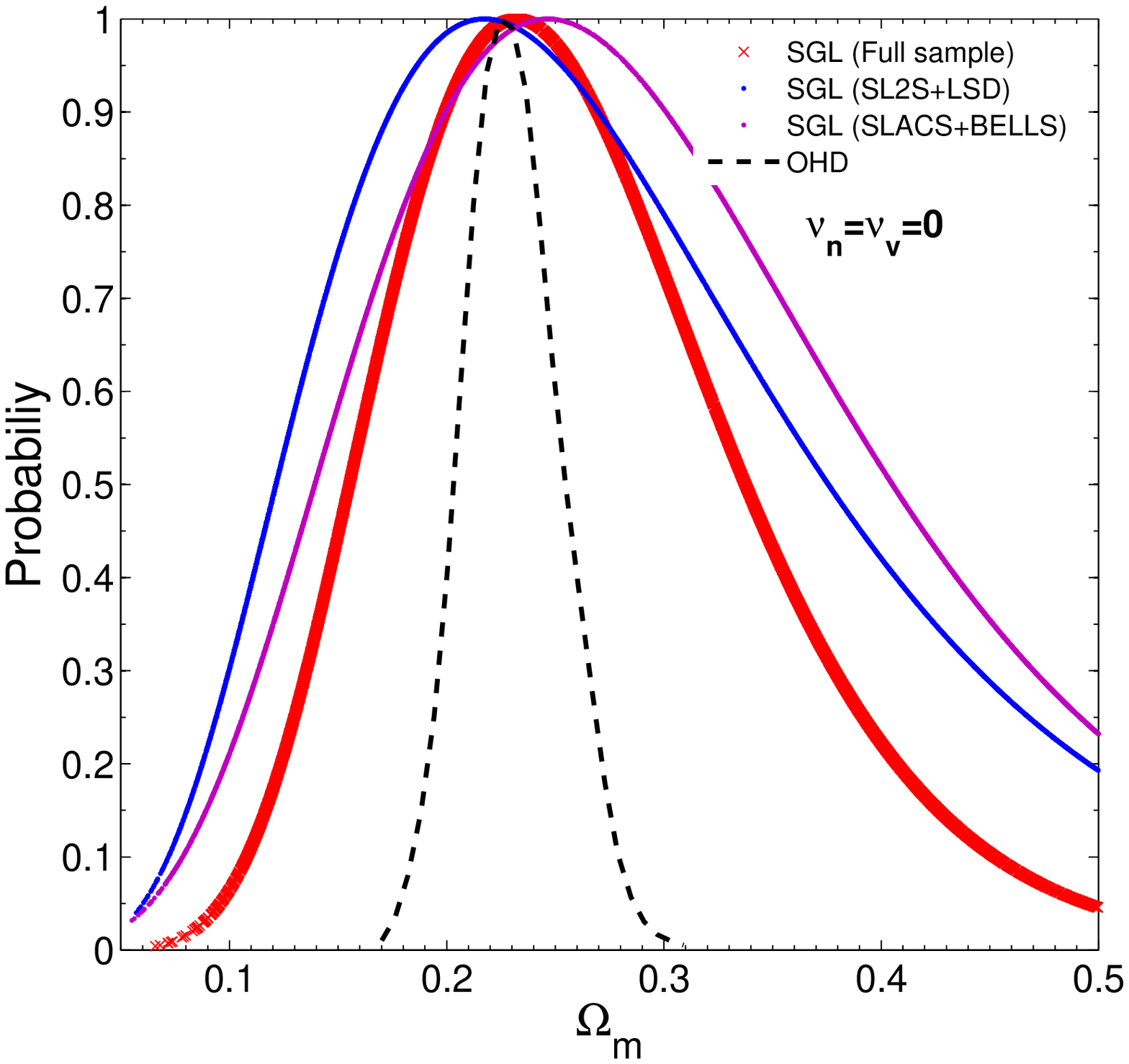}  \includegraphics[width=0.45\textwidth]{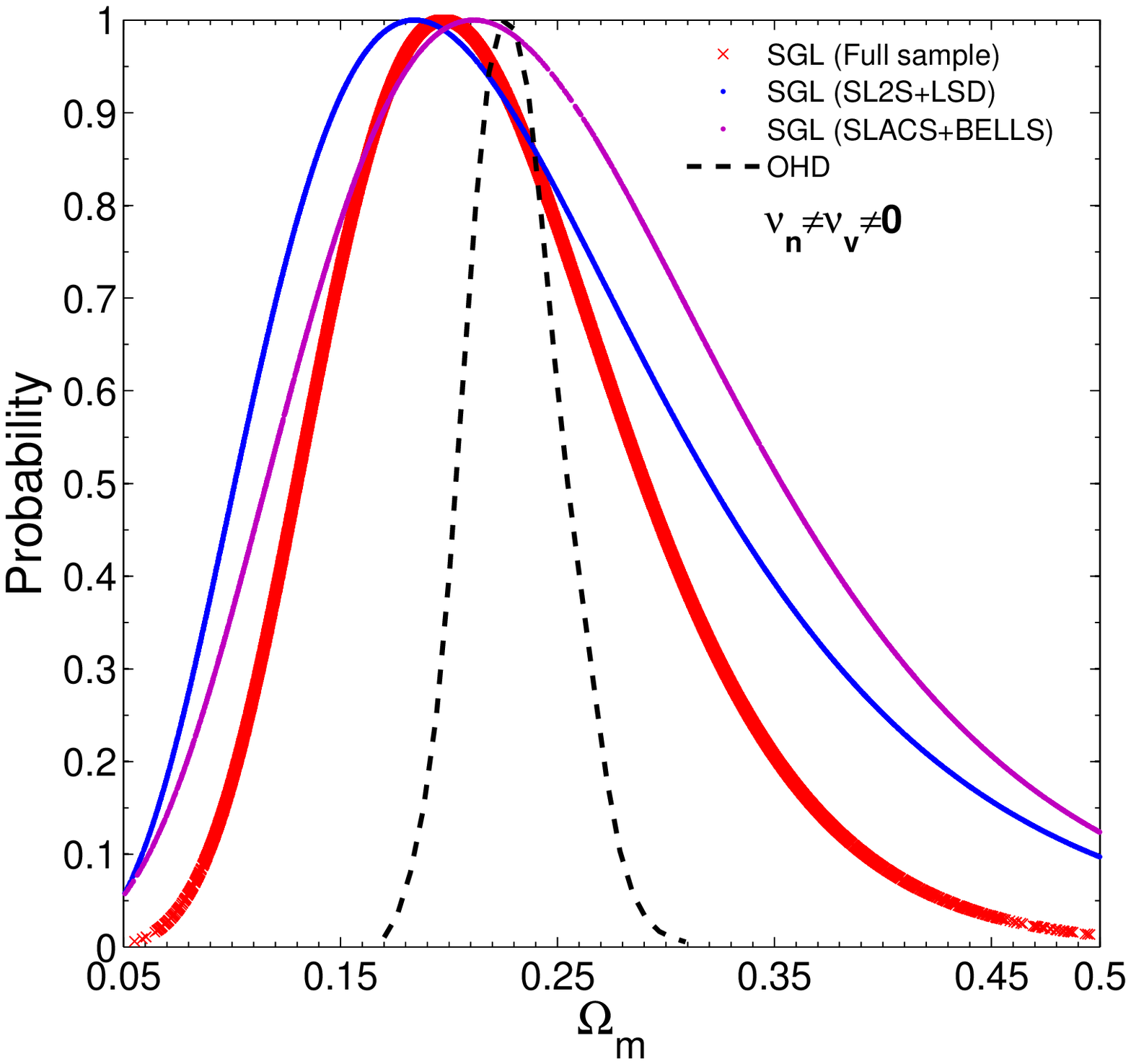}
\caption{ Constraints on the matter density parameter in the flat
DGP model, which are obtained from the lens redshift distribution of
current SGL systems with and without the redshift evolution of
lensing galaxies. Fitting results from recent observational Hubble
parameter data (OHD) are also added for comparison.} \label{fig3}
\end{figure*}

\begin{table*}
\begin{center}
\caption{\label{tab:result} Summary of the cosmological constraints
from the lens redshift distribution of current strong lensing
observations.}
\begin{tabular}{c|l|l}\hline\hline
Cosmological model   & Data  & Cosmological fit  \\
\hline
$\Lambda$CDM ($\nu_n=\nu_v=0$) & Current SGL (Full sample) & $\Omega_m=0.315\pm0.085$      \\
~~~~~~~~~~~ ($\nu_n=\nu_v=0$) & Current SGL (Sample A) & $\Omega_m=0.291\pm0.109$      \\
~~~~~~~~~~~ ($\nu_n=\nu_v=0$) & Current SGL (Sample B) & $\Omega_m=0.355\pm0.125$      \\

 ~~~~~~~~~~~($\nu_n\neq \nu_v\neq0$) & Current SGL (Full sample)  & $\Omega_m=0.274\pm0.076$       \\
 ~~~~~~~~~~~ ($\nu_n\neq \nu_v\neq0$) & Current SGL (Sample A) & $\Omega_m=0.254\pm0.096$      \\
 ~~~~~~~~~~~ ($\nu_n\neq \nu_v\neq0$) & Current SGL (Sample B) & $\Omega_m=0.314\pm0.116$      \\

  \hline

 DGP ($\nu_n=\nu_v=0$) & Current SGL (Full sample) & $\Omega_m=0.243\pm0.077$      \\

 ~~~~~~~~~~~ ($\nu_n=\nu_v=0$) & Current SGL (Sample A) & $\Omega_m=0.238\pm0.105$      \\
~~~~~~~~~~~ ($\nu_n=\nu_v=0$) & Current SGL (Sample B) & $\Omega_m=0.263\pm0.111$      \\
   ~~~~~~~~~~~($\nu_n\neq \nu_v\neq0$) & Current SGL (Full sample)  & $\Omega_m=0.207\pm0.067$       \\
~~~~~~~~~~~ ($\nu_n\neq \nu_v\neq0$) & Current SGL (Sample A) & $\Omega_m=0.204\pm0.092$      \\
  ~~~~~~~~~~~ ($\nu_n\neq \nu_v\neq0$) & Current SGL (Sample B) & $\Omega_m=0.228\pm0.101$      \\

\hline\hline
\end{tabular}
\end{center}
\end{table*}

\subsection{The standard cosmological model ($\Lambda$CDM)}

The cosmological model containing a cosmological constant and cold
dark matter (CDM) component is usually called the standard
cosmological model. The unique feature of the cosmological constant
is that its equation-of-state parameter $w=-1$. Therefore, assuming
a flat Universe with negligible radiation density
($\Omega_m+\Omega_\Lambda=1$), ordinary pressureless dust matter and
the cosmological constant contribute to the total energy. The
Friedmann equation is
\begin{equation}
H^2 = H_0^2 \ [\Omega_m(1+z)^3+1-\Omega_m] \, ,
\end{equation}
where $\Omega_m$ parameterizes the density of matter (both baryonic
and non-baryonic components) in the Universe. Therefore, if spatial
flatness of the FRW metric is assumed, this model has only one
independent parameter ($\Omega_m$).

By fitting the $\Lambda$CDM model to the current 158 strong lensing
systems, we get $\Omega_m=0.315\pm0.085$ in the case of no evolution
model ($\nu_n=\nu_v=0$), which is well consistent with the results
given by the recent data release of \textit{Planck} observations
\citep{Planck15}. One can clearly see that the currently compiled
strong lensing data improves the constraints on model parameters
significantly. Considering Sample A and Sample B, the likelihood is
maximized at $\Omega_m=0.291\pm0.109$ and $\Omega_m=0.355\pm0.125$
with no redshift evolution. More importantly, we find that different
galaxy evolution models will slightly affect the constraints on the
model parameter: the evolution of the quantities $n_{*}$ and
$\sigma_{*}$ will shift the the matter density parameter to a lower
value. For the three strong lensing samples defined in Section 2,
the best-fitted values and the 1$\sigma$ limits are
$\Omega_m=0.274\pm0.076$ (Full sample), $\Omega_m=0.254\pm0.096$
(Sample A), and $\Omega_m=0.314\pm0.116$ (Sample B) using the
power-law evolution model after \citet{Kang05}. These results are
shown in Fig.~2 and Table 1. We remark here that our results
strongly suggest that larger and more accurate sample of the strong
lensing data can become an important complementary probe to test the
the properties of dark energy. This conclusion is strengthened by
the comparison of our cosmological fits from the redshift
distribution of a larger sample and those from the absolute lensing
probability for a smaller sample of optical and radio lenses
($\Omega_m=0.3^{+0.2}_{-0.1}$) \citep{Chiba99}.

Another important issue is the comparison of our cosmological
results with earlier studies done using other alternative probes. We
turn to the observational Hubble parameter data (OHD) to verify this
point. The Hubble parameter $H(z)$ at 31 different redshifts was
obtained from the differential ages of passively evolving galaxies,
while 10 more Hubble parameter data were determined recently from
the radial BAO size method (see \citet{Qi18} for more details). With
the latest OHD data comprising 41 data points, we obtain the
best-fit values of the cosmological parameters in the flat
$\Lambda$CDM model: ${\Omega_m}=0.255\pm0.030$ and $H_0=70.4\pm2.5
\; \rm{kms}^{-1} \; \rm{Mpc}^{-1}$ at 68.3\% confidence level. For a
good comparison, fits on the matter density parameter are also
plotted in Fig.~2 (with Hubble constant marginalized). One may
observe that the results obtained from the lens redshift test are
well consistent with the OHD fits, although larger uncertainties may
arise due to possible evolution of the quantities $n_{*}$ and
$\sigma_{*}$. Such excellent consistency could also be clearly seen
through the comparison with WMAP 5-year data combined with BAO and
SN Union data sets \citep{Komatsu09}, in which the best-fit
parameters are given as $\Omega_m=0.274$ and $H_0=70.5 \rm{kms}^{-1}
\; \rm{Mpc}^{-1}$ for the flat $\Lambda$CDM model. In contrast,
recent CMB anisotropy measurements by \textit{Planck} data favors a
higher value of $\Omega_m$ and thus a larger matter density in the
$\Lambda$CDM model. Based on the full-mission \textit{Planck}
observations of temperature and polarization anisotropies of the CMB
radiation, \textit{Planck} Collaboration (2015) gave the best-fit
parameter: ${\Omega_m}=0.308\pm0.012$ and $H_0=67.8\pm0.9 \;
\rm{kms}^{-1} \; \rm{Mpc}^{-1}$ \citep{Planck15}. Let us note that
the matter density parameter inferred from CMB and OHD data are
highly dependent on the value of the Hubble constant, considering
the well known strong degeneracy between $\Omega_m$ and $H_0$.
Therefore independent measurement of $\Omega_m$ from strong lensing
statistics could be expected and indeed is revealed here.

\subsection{Dvali-Gabadadze-Porrati model (DGP)}

This DGP model is one of the simplest modified gravity models based
on the concept of brane world theory, in which gravity leaks out
into the bulk above a certain cosmological scale $r_c$. This
provides a mechanism for the accelerated expansion without
introducing a repulsive-gravity fluid \citep{Dvali00}. In the
framework of a spatially flat DGP model, the Friedmann equation is
modified as
\begin{equation}
H^2-{H\over r_c}= \frac{8\pi G}{3} \rho_m \, ,
\end{equation}
where $r_c=(H_0(1-\Omega_m))^{-1}$ is the length at which the
leaking occurs. The above equation can be directly rewritten to
generate the expansion rate
\begin{equation}
H^2 = H_0^2
(\sqrt{\Omega_{m}(1+z)^3+\Omega_{r_c}}+\sqrt{\Omega_{r_c}})^2 \, ,
\end{equation}
where an adimensional parameter is associated with the cosmological
scale through $\Omega_{rc}=1/(4r_c^2H_0^2)$. It is straightforward
to check the validity of the relation $\Omega_{r_c} = \frac{1}{4} (1
- \Omega_m)^2$ in the flat DGP model, which indicates that there is
only one free parameter in this model ($\Omega_m$).

Working on the DGP model, we obtain the fitting results from two
cases of evolution models of lensing galaxies, which are displayed
in Fig.~3 and Table 1. The marginalized 1$\sigma$ constraints of the
parameters are: $\Omega_m=0.243\pm0.077$ with no redshift evolution
and $\Omega_m=0.207\pm0.067$ with redshift evolution. In both cases,
the strong lensing statistics imposes a strong bound on $\Omega_m$,
which is similar to what was obtained when the dark energy models
are explored with the ratio of (angular-diameter) distances between
lens and source and between observer and lens \citep{Cao12b,Cao15}.
Working on the two sub-samples, the best-fit values of the
parameters are: $\Omega_m=0.238\pm0.105$ (with no redshift
evolution), $\Omega_m=0.204\pm0.092$ (with redshift evolution) for
Sample A, and $\Omega_m=0.263\pm0.111$ (with no redshift evolution),
$\Omega_m=0.228\pm0.101$ (with redshift evolution) for Sample B.
More interestingly, we also note the DGP model, which has already
been ruled out observationally considering the precision
cosmological observational data \citep{Fang08,Durrer10,Maartens10},
seems to be a representative set instead of viable candidates for
dark energy. Such tendency is also strongly hinted by the fitting
results derived from the lens redshift test and the latest Hubble
parameter data (see Fig.~3).

Now it is worthwhile to make some comments on the results obtained
above. Firstly, comparing to the previous analysis with a smaller
sample \citep{Cao12a,Cao12b}, our results strongly suggest that
larger and more accurate sample of SGL data can become an important
complementary probe to other standard ruler data. More importantly,
the advantage of our method lies in the benefit of being independent
of the Hubble constant. Consequently, $H_0$ and its uncertainty do
not influence the final cosmological results. Secondly, in the
framework of two cosmologies classified into different categories,
the null hypothesis of a dominant matter density ($\Omega_m\sim 1$)
is excluded at large confidence level ($>4\sigma$). Therefore, our
results has provided independent evidence for the accelerated
expansion of the Universe, which is the most unambiguous result of
the current dataset. Thirdly, considering the general concern that
strong gravitational lenses could be a biased sample of galaxies, we
note that systematic errors due to sample incompleteness do not
exceed $\sim 0.1$ on the matter density parameter. Finally, although
constraints on the hierarchical models of galaxy evolution is beyond
the scope of this work, simple evolution of the velocity dispersion
function does not significantly affect the lensing statistics and
thus the derivation of cosmological information. This conclusion
agrees very well with the previous studies on lensing statistics of
early-type galaxies \citep{Cha03,Mit05,CN07}.

\begin{figure}
   \centering
   \includegraphics[width=0.5\textwidth]{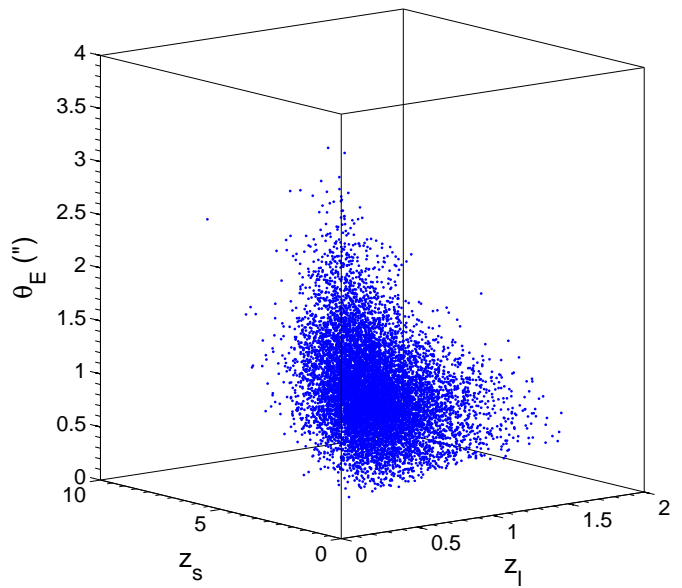}
\caption{ Scatter plot of 10000 simulated strong lensing systems
from future LSST surveys.} \label{fig4}
\end{figure}

\begin{figure}
\begin{center}
\includegraphics[width=1.0\hsize]{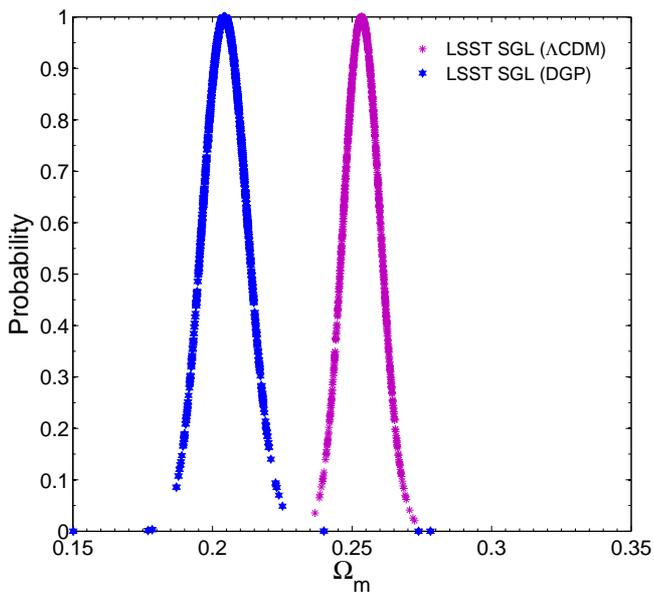}
\end{center}
\caption{ Constraints on the cosmological parameters from the
simulated LSST strong lensing data. \label{fig5}}
\end{figure}

\subsection{Cosmology from future LSST observations}

The lensing constraints on the cosmological parameters are already
quite competitive compared with those from other methods. However,
they still suffer from the small number of lenses in our statistical
sample. The redshift distribution test, with larger gravitational
lensing samples from future wide-field surveys, could be helpful for
advancing such applications. Following the recent analysis
\citep{Kuhlen04,Marshall05}, benefit from the improved depth, area
and resolution, the next generation wide and deep sky surveys will
increase the current galactic-scale lens sample sizes by orders of
magnitude in the near future. Recent analytical work has forecast
the number of galactic-scale lenses to be discovered in the
forthcoming photometric surveys \citep{Collett15}. With a large
increase to the known strong lens population, current work could be
extended to a new regime: in the framework of lens redshift test,
what kind of cosmological results one could obtain from $\sim 10000$
discoverable lens population in the forthcoming Large Synoptic
Survey Telescope (LSST) survey.

Using the simulation programs publicly available \footnote
{github.com/tcollett/LensPop}, we obtain 10000 strong lensing
systems on the base of realistic population models of elliptical
galaxies acting as lenses, whose mass distribution is approximated
by the singular isothermal ellipsoids. Following the assumptions
underlying the simulation, we take the VDF of elliptical galaxies in
the local Universe derived from the SDSS Data Release 5
\citep{Choi07}. Meanwhile, in our simulation we assume that neither
the shape nor the normalization of this function vary with redshift,
which is well consistent with the previous studies on lensing
statistics \citep{Cha03,Mit05,CN07} and the recent observations of
\citet{Bezanson11}. Fig.~4 shows the scatter plot of the simulated
lensing systems, from which one can see the LSST lenses resulted in
a fair coverage of lenses and sources redshifts.

Then we assess the likelihood $\mathcal{L}$ of the observed lens
redshift from the strong lensing data, with the results summarized
in Fig.~5. The effectiveness of of our method could be seen from the
discussion of this question: \emph{Is it possible to achieve a
stringent measurement of the present value of the matter density
parameter?} As is clearly shown in Fig.~5, in the framework of two
different cosmological models, one can expect the matter density
parameter to be estimated with the precision of $\Delta\Omega_m\sim
0.006$. Therefore, with about 10000 discoverable SGL systems in
forthcoming surveys, the lens redshift test places more stringent
constraints on the matter density parameter, compared with the
combined results from \textit{Planck} temperature and lensing data
($\Delta\Omega_m=0.012$) \citep{Planck15}. Such conclusion could
also be obtained from the comparison between our results and those
using the future baryon acoustic oscillation (BAO) and supernova
observations of the Joint Dark Energy Mission (JDEM) project in the
low redshift region \citep{Zhao11}. Therefore, we have added some
support to the argument that with more detectable galactic-scale
lenses from the forthcoming surveys, the lens redshift distribution
can eventually be used to carry out stringent tests on various
cosmological models. However, one should note that sample
incompleteness still could constitute an important source of
systematic errors in the future, i.e., the current systematics of
$\sim$0.1 might dominate over the statistical uncertainty of the
matter density parameter. Therefore, in order to improve constraints
on cosmological parameters, our findings strongly motivate the
future use of a larger sample of gravitational lenses in the
forthcoming surveys, for which completeness is homogenous as a
function of the lensed image separation and the lens redshift
\citep{CN07}.

\begin{table*}
\begin{center}
\caption{\label{tab:result} Summary of the lensing based
characteristic velocity dispersion and its corresponding ratio to
the stellar velocity dispersion, based on the lens redshift
distribution of current strong lensing observations. }
\begin{tabular}{c|l|l|l}\hline\hline
Galaxy evolution model & Data & Lening based velocity dispersion (km/s) & Ratio  \\
\hline
$\nu_n=\nu_v=0$ & Current SGL (Full sample) & $\sigma_{*,lens}=219.1\pm5.5$ & $f_E=1.010\pm0.025$     \\
$\nu_n=\nu_v=0$ & Current SGL (Sample A) & $\sigma_{*,lens}=224.5\pm11.3$ & $f_E=1.034\pm0.052$      \\
  $\nu_n=\nu_v=0$ & Current SGL (Sample B) & $\sigma_{*,lens}=217.3\pm6.3$ & $f_E=1.001\pm0.029$      \\

$\nu_n\neq \nu_v\neq0$ & Current SGL (Full sample)  & $\sigma_{*,lens}=221.6\pm5.6$ & $f_E=1.021\pm0.026$       \\
$\nu_n\neq \nu_v\neq0$ & Current SGL (Sample A) & $\sigma_{*,lens}=228.3\pm11.7$ & $f_E=1.052\pm0.054$      \\
  $\nu_n\neq \nu_v\neq0$ & Current SGL (Sample B) & $\sigma_{*,lens}=219.4\pm6.4$ & $f_E=1.011\pm0.029$      \\

\hline\hline
\end{tabular}
\end{center}
\end{table*}

\begin{figure*}
   \centering
   \includegraphics[width=0.45\textwidth]{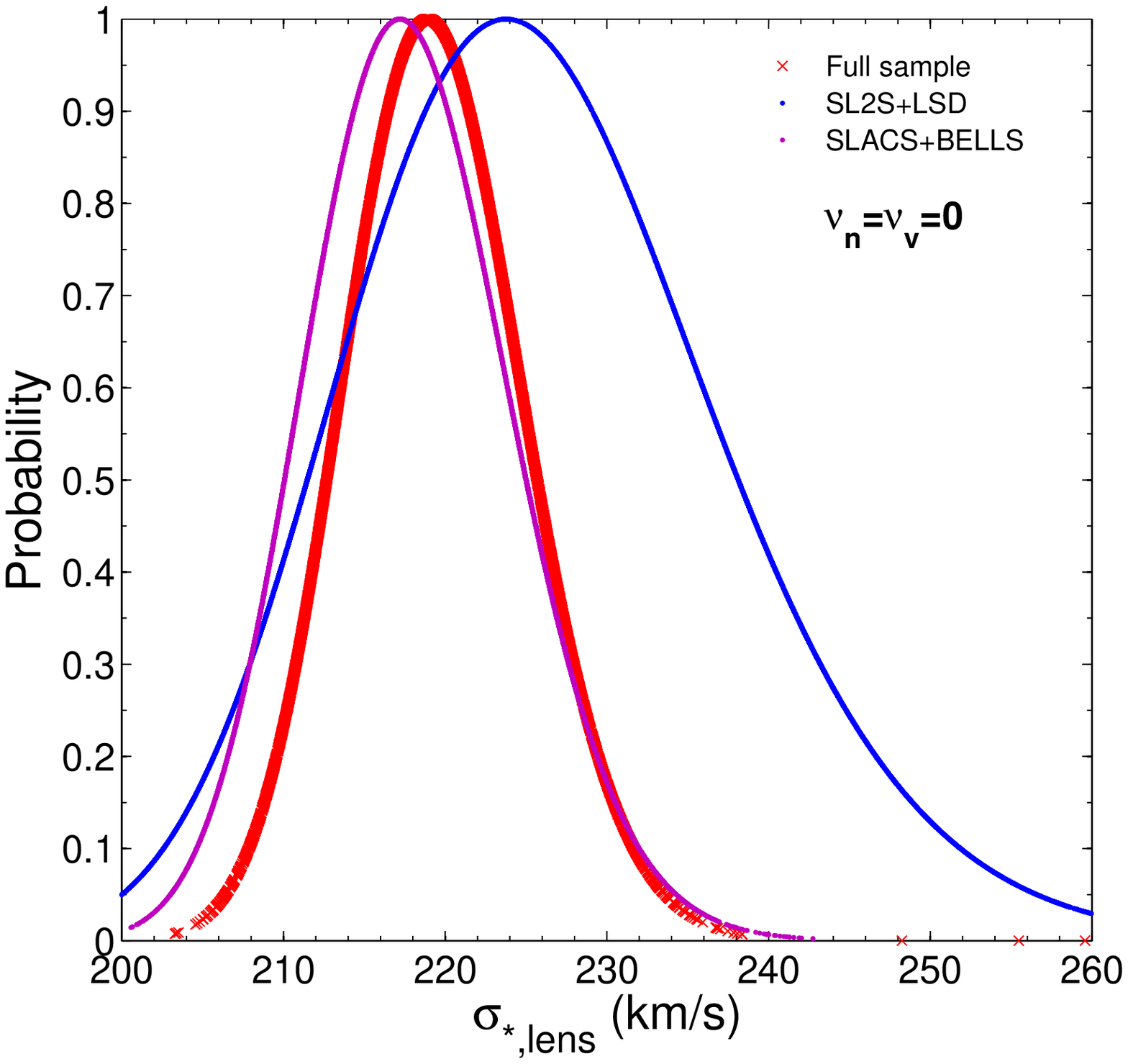} \includegraphics[width=0.45\textwidth]{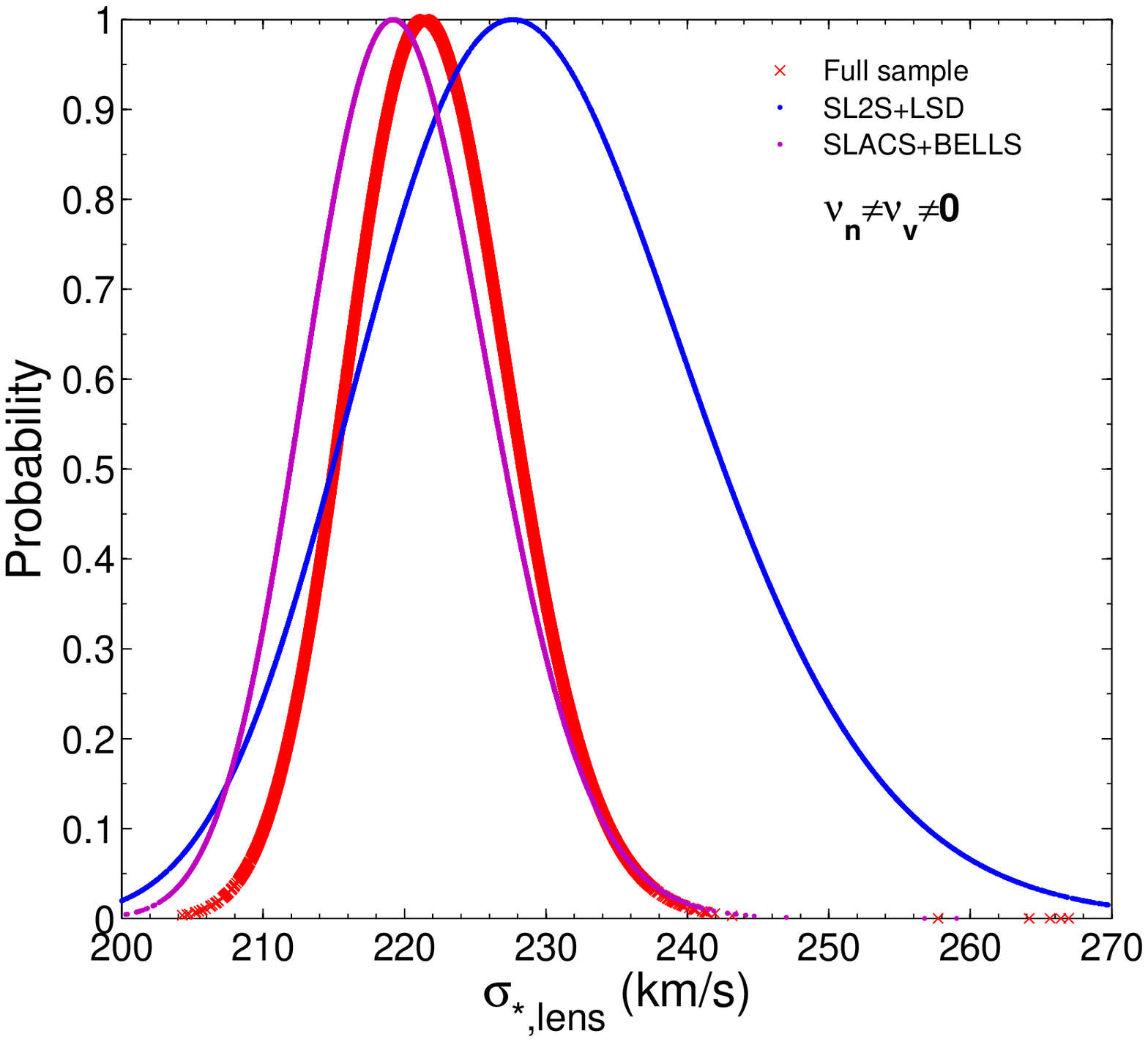}
\caption{Confidence limits on the lensing based characteristic
velocity dispersion of early-type galaxies.} \label{fig6}
\end{figure*}

\begin{figure*}
   \centering
   \includegraphics[width=0.45\textwidth]{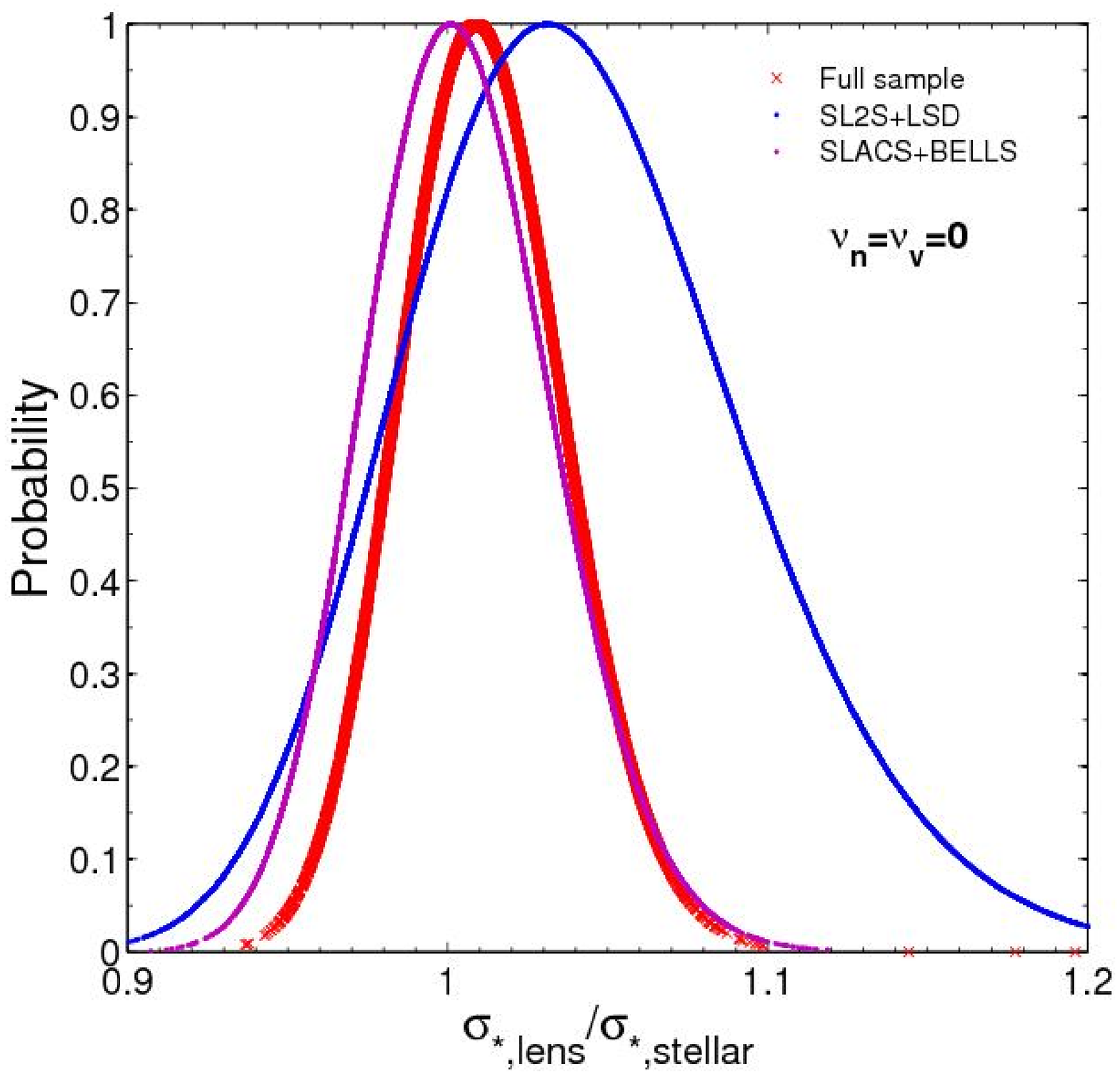} \includegraphics[width=0.45\textwidth]{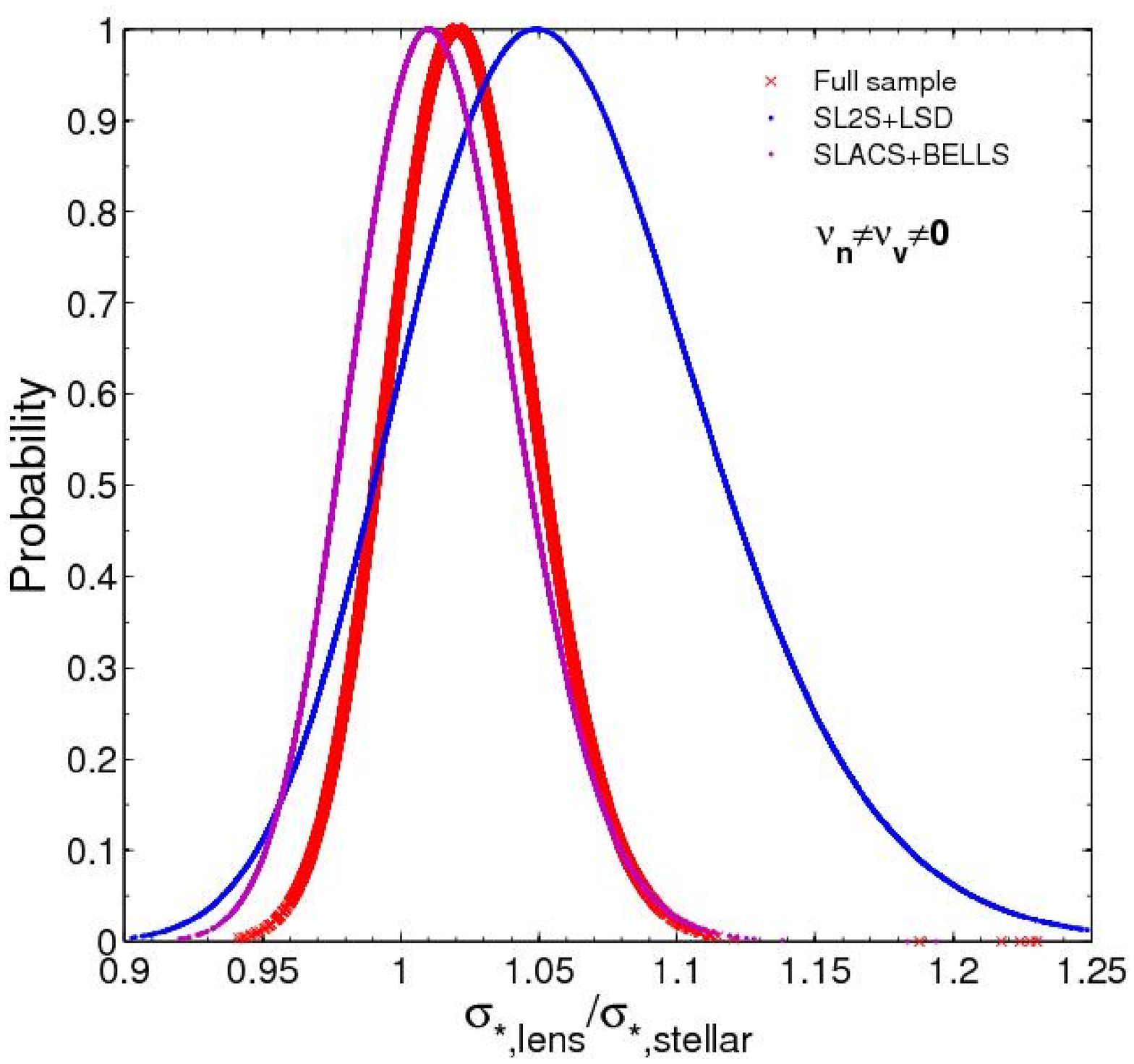}
\caption{Confidence limits on the ratio of the lensing based
velocity dispersion to the stellar velocity dispersion.}
\label{fig7}
\end{figure*}

\section{Constraints on lensing based characteristic velocity dispersion}

The velocity dispersion functions (VDF) of early-type galaxies,
which can be inferred from early-type luminosity functions via an
adopted power-law relation between luminosity and velocity
dispersion (the Faber-Jackson relation), are crucial observables to
provide powerful constraints on predictions of models of galaxy
formation and evolution. On the side of the measurement of VDF, the
first direct measurement of the VDF of early-type galaxies was made
by \citet{Sheth03}, based on Sloan Digital Sky Survey (SDSS) DR1 of
9000 early-type galaxies \citep{Bernardi03}. Then \citet{Choi07}
obtained a new VDF based on the much larger SDSS DR5, which is quite
different from the DR1 VDF in the characteristic velocity dispersion
at 1$\sigma$. A possible inconsistency between the two VDF
measurements can be largely attributed to the improved galaxy
classification scheme, making use of a SDSS $u-r$ color versus $g-i$
color gradient space \citep{Park05}.

In this work, we consider constraining a model VDF of early-type
galaxies using the statistics of strong gravitational lensing. More
specifically, the distribution of lens redshift is mainly applied to
place limits on the characteristic velocity dispersion. Moreover,
considering the strong degeneracy between the shape of the VDF
($\alpha$, $\beta$) and the characteristic velocity dispersion
($\sigma_*$) \citep{Chae05}, the focus of this work is: What would
be the constrained value of $\sigma_*$ if $\alpha$ and $\beta$ are
fixed by a stellar VDF? We obtain a solely lensing-based VDF
assuming no and passive evolution of early-type galaxies, which will
then be compared with the measured VDF in the local universe. Fig.~6
shows the fits on $\sigma_*$ for the case of fixing $\alpha$ and
$\beta$ by the type-specific VDF \citep{Chae10}:
$\sigma_{*,lens}=219.1\pm5.5$ km/s (with no redshift evolution),
$\sigma_{*,lens}=221.6\pm5.6$ km/s (with redshift evolution) for the
full sample, $\sigma_{*,lens}=224.5\pm11.3$ km/s (with no redshift
evolution), $\sigma_{*,lens}=228.3\pm11.7$ km/s (with redshift
evolution) for Sample A, and $\sigma_{*,lens}=217.3\pm6.3$ km/s
(with no redshift evolution), $\sigma_{*,lens}=219.4\pm6.4$ km/s
(with redshift evolution) for Sample B. Our results demonstrate the
strong consistency between the lensing-based value of
$\sigma_{*,lens}$ and the corresponding stellar values for the
adopted stellar VDF, which, to some extent agrees with the velocity
dispersion profiles of a sample of 37 elliptical galaxies using a
Jaffe stellar density profile and the SIS model for the total mass
distribution \citep{Kochanek94}.

Let us note here that the velocity dispersion $\sigma_{*,lens}$ of
the mass distribution and the observed stellar velocity dispersion
$\sigma_{*,stellar}$ need not be the same. We adopt a parameter
$f_E=\sigma_{*,lens}/\sigma_{*,stellar}$ that relates the velocity
dispersion and the spectroscopically measured central stellar
dispersion. Based on the three different strong lensing samples, we
obtain the following best-fitting values and corresponding 68\%
confidence level uncertainties: $f_E=1.010\pm0.025$ (with no
redshift evolution), $f_E=1.021\pm0.026$ (with redshift evolution)
for the full sample, $f_E=1.034\pm0.052$ (with no redshift
evolution), $f_E=1.052\pm0.054$ (with redshift evolution) for Sample
A, and $f_E=1.001\pm0.029$ (with no redshift evolution),
$f_E=1.011\pm0.029$ (with redshift evolution) for Sample B. It is
apparent that for each case, the consistency between the velocity
dispersion for our power-law lens model and the spectroscopically
measured central stellar dispersion is supported within $1\sigma$
C.L. However, the constrained results on $f_E$ parameter are still
particularly interesting. What would be an appropriate
interpretation of such possible disagreement between
$\sigma_{*,lens}$ and $\sigma_{*,stellar}$? Note that the real
early-type galaxies can be divided into the luminous stellar
component and the extended dark matter halo component. Based on the
X-ray properties of the first X-ray-complete optically selected
sample of elliptical galaxies, \citet{White96} discussed the kinetic
temperature of the gas and the stars. The derived results and other
independent results \citep{White98,Loewenstein99} indicate that dark
matter halos are dynamically hotter than the luminous stars, which
strongly implies a greater velocity dispersion of dark matter than
the visible stars. More recently, \citet{Treu04} used a sample of
five individual lens systems to determine the ratio of the SIE
velocity dispersion to the stellar velocity dispersion, producing a
mean value of $f_E=1.15\pm0.05$ from optical spectroscopic
observation of the lensing galaxies. Therefore, our results
presented in Fig.~7 and Table 2 robustly indicate the possible
presence of dark matter, in the form of a mass component with
velocity dispersion greater than stellar velocity dispersion.

Finally, we illustrate what kind of result could be obtained from
the future data in the forthcoming LSST survey. The resulting
constraint on the $f_E$ parameter becomes $\Delta f_E=0.003$, with
the posterior probability density shown in Fig.~8. It can be clearly
seen that much more stringent constraints would be achieved, and one
can expect $f_E$ to be estimated with $10^{-3}$ precision.
Therefore, the lens redshift test, when applied to larger samples of
strong lensing systems, can provide an independent and alternative
experiment to test the global properties of early-type galaxies at
much higher accuracy.

\begin{figure}
   \centering
\includegraphics[width=0.5\textwidth]{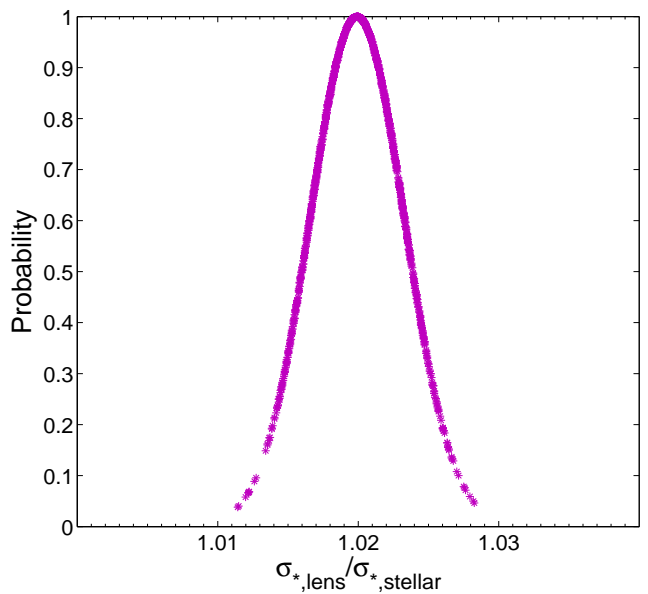}
\caption{Confidence limits on the ratio of the SIE velocity
dispersion to the stellar velocity dispersion, which are derived
from the simulated LSST strong lensing data.} \label{fig8}
\end{figure}

\begin{figure}
\begin{center}
\includegraphics[width=1.0\hsize]{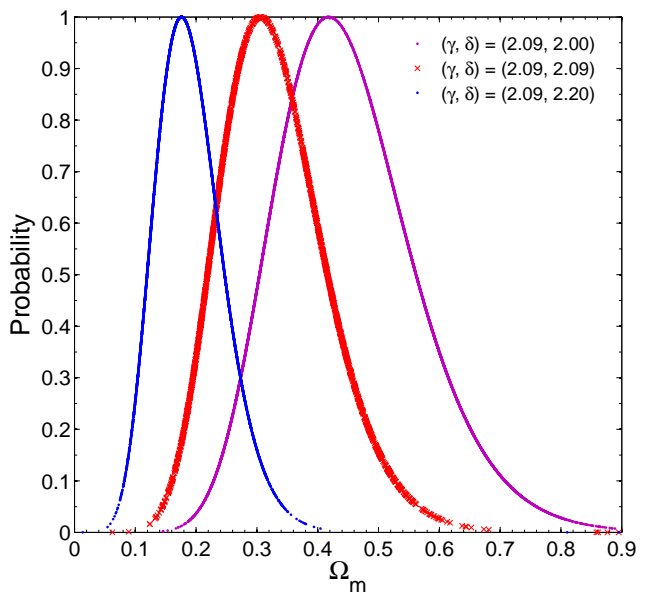}
\end{center}
\caption{ Constraints on the matter density parameter (in the flat
$\Lambda$CDM model) from the current SGL systems, with different
luminosity density profiles for the lensing galaxies ($\propto
r^{-\delta}$). \label{fig9}}
\end{figure}

\section{Conclusion and discussion}
\label{sec:conclusion}

In this work, based on a well-defined sample of lensing, elliptical
galaxies drawn from a large catalog of 158 gravitational lenses, we
use the statistical properties of the strong lens sample (i.e., the
redshift distribution of lenses) to constrain the cosmological
parameters and the velocity dispersion functions (VDF) of early-type
galaxies. In order to assess the accuracy of the results, two cases
of VDF evolution models of lensing galaxies are taken into account.
Moreover, we have quantified the ability of future measurements of
strong lensing systems from the forthcoming Large Synoptic Survey
Telescope (LSST) survey, which encourages us to probe cosmological
parameters and early-type galaxy properties at much higher accuracy.
Here we summarize our main conclusions in more detail:

\begin{itemize}

\item Firstly of all, with the current catalog of 158 gravitational
lenses, we evaluate the power of direct measurements of lens
redshift distribution on constraining two popular cosmological
models. For the concordance $\Lambda$CDM model, we have found
$\Omega_m=0.315\pm0.085$ with no redshift evolution and
$\Omega_m=0.274\pm0.076$ with redshift evolution. For the DGP
brane-world scenario, the current strong lensing systems provide the
constraints on the matter density parameter as
$\Omega_m=0.243\pm0.077$ with no redshift evolution and
$\Omega_m=0.207\pm0.067$ with redshift evolution. More importantly,
the DGP model, which has already been ruled out observationally
considering the precision cosmological observational data, seems to
be a representative set instead of viable candidates for dark
energy. Two additional sub-samples are also included to account for
possible selection effect introduced by the detectability of lens
galaxies, which confirms that systematic errors due to sample
selection are not larger than statistical uncertainties. Whereas,
there are several sources of systematics we do not consider in this
paper. For instance, although the average total power-law density
slope of observed early-type galaxies has been found to be close to
isothermal within a few effective radii \citep{Wang18}, the scatter
of other galaxy structure parameters, especially those
characterizing the stellar distribution in the lensing galaxies,
could be an important source of systematic errors on the final
results. An influential paper by \citet{Hernquist90} suggested that
a brand new Hernquist profile can provide a good approximation to
the luminosity distribution of spherical galaxies. Such density
profile, which resembles an elliptical galaxy or dark matter halo
with $r^{-1}$ at small radii and $r^{-4}$ at large radii, has found
widespread astrophysical applications in the literature
\citep{Barnes92,Lowing11,Ngan15}. Therefore, we perform a
sensitivity analysis to investigate how the cosmological constraint
on flat $\Lambda$CDM is altered by the luminosity density profile.
In the framework of a general mass model for the total-mass density
and luminosity density (Eq.~(\ref{eq:rhopl})), the
luminosity-density slope is varying as $\delta=$2.00, 2.09, and
2.20, while total-mass density parameter is fixed at its best-fit
value ($\gamma=2.09$) from the total-mass and stellar-velocity
dispersion measurements of a sample of SLACS lenses
\citep{Koopmans09}. In general, one can see from Fig.~9 that the
derived value of $\Omega_m$ is sensitive to the adopted luminosity
density profiles, i.e., a steeper stellar density profile in the
early-type galaxies will shift the matter density parameter to a
relatively lower value. This illustrates the importance of using
auxiliary data to improve constraints on the luminosity density
parameter, with future high-quality integral field unit (IFU) data
\citep{Barnabe13}.

\item The advantage of our method lies in the benefit of being
independent of the Hubble constant. Therefore independent
measurement of $\Omega_m$ from strong lensing statistics could be
expected and indeed is revealed here. More interestingly, one may
also observe that simple VDF evolution does not significantly affect
the lensing statistics and thus the derivation of cosmological
information, if all galaxies are of early type. In the framework of
two cosmologies classified into different categories, the null
hypothesis of a vanishing dark energy density is excluded at large
confidence level ($>4\sigma$). Therefore, our results has provided
independent evidence for the accelerated expansion of the Universe,
which is the most unambiguous result of the current dataset.

\item Moreover, we have quantified the ability of a future
measurements of SGL from the forthcoming LSST survey, which may
detect tens of thousands of lenses for the most optimistic scenario
\citep{Collett15}. In the framework of the two cosmological models,
one can expect $\Omega_m$ to be estimated with the precision of
$\Delta\Omega_m\sim 0.006$. Therefore, with about 10000 discoverable
SGL systems in forthcoming surveys, the lens redshift test places
more stringent constraints on the matter density parameter, compared
with the combined results from \textit{Planck} temperature and
lensing data ($\Delta\Omega_m=0.012$) \citep{Planck15}. Therefore,
we have added some support to the argument that the lens redshift
distribution, with more detectable galactic-scale lenses from the
forthcoming surveys, can eventually be used to carry out stringent
tests on various cosmological models.

\item Finally, the currently available lens redshift distribution,
which constitutes a promising new cosmic tracer, may also allow us
to obtain stringent constraints on the global properties of
early-type galaxies. We use mainly the distribution of lens redshift
to constrain the characteristic velocity dispersion (with fixed
shape of the velocity function), and thus obtain a solely
lensing-based VDF for $z_l\sim1.0$. Our results demonstrate the
strong consistency between the lensing-based value of
$\sigma_{*,SIE}$ and the corresponding stellar value
$\sigma_{*,stellar}$ for the adopted stellar VDF in the local
universe. Furthermore, a parameter
$f_E=\sigma_{*,SIE}/\sigma_{*,stellar}$ is adopted to quantify the
relation between the two velocity dispersions, which is fit to
$f_E=1.010\pm0.025$ (with no redshift evolution) and
$f_E=1.021\pm0.026$ (with redshift evolution) from the full SGL
sample. Therefore, our results agrees with the respective values of
$f_E$ derived in the previous studies, which robustly indicates the
possible presence of dark matter halos in the early-type galaxies,
with velocity dispersion greater than stellar velocity dispersion.
More importantly, this statistical lensing formalism, when applied
to larger samples of strong lensing systems, can provide much more
stringent constraints and one can expect $f_E$ to be estimated with
$10^{-3}$ precision.

\end{itemize}

\section*{Acknowledgments}

This work was supported by National Key R\&D Program of China No.
2017YFA0402600; the National Natural Science Foundation of China
under Grants Nos. 11705107, 11475108, 11503001, 11373014, 11073005
and 11690023; Beijing Talents Fund of Organization Department of
Beijing Municipal Committee of the CPC; and the Opening Project of
Key Laboratory of Computational Astrophysics, National Astronomical
Observatories, Chinese Academy of Sciences. Y. Pan was supported by
the Scientific and Technological Research Program of Chongqing
Municipal Education Commission (Grant No KJ1500414); and Chongqing
Municipal Science and Technology Commission Fund
(cstc2015jcyjA00044, and cstc2018jcyjAX0192).

%

\end{document}